# Breaking Classical Public Key Cryptosystems by Using a Novel Ensemble Search Algorithm

Chien-Yuan Chen
Department of Information Engineering,
I-Shou University, Kaohsiung County, Taiwan, 84008 R.O.C.
E-mail: cychen@isu.edu.tw

## Abstract

In this paper, we improve Bruschweiler's algorithm such that only one query is needed for searching the single object z from $N=2^n$ unsorted elements. Our algorithm construct the new oracle query function $g(\cdot)$ satisfying $g(x)=0$ for all input x, except for one, say $x=z$, where $g(z)=z$. To store z, our algorithm extends from one ancillary qubit to n ancillary qubits. We then measure these ancillary qubits to discover z. We further use our ensemble search algorithm to attack classical public key cryptosystems. Given the ciphertext $C=E_k(m, r)$ which is generated by the encryption function $E_k()$, a public key k, a message m, and a random number r, we can construct an oracle query function $h(\cdot)$ satisfying $h(m', r')=0$ if $E_k(m', r') \neq C$ and $h(m', r')= (m', r')$ if $E_k(m', r')=C$. There is only one object, say (m, r), can be discovered in decryption of C. By preparing the input with all possible states of (m', r'), we can thus use our ensemble search algorithm to find the wanted object (m, r). Obviously, we break the classical public key cryptosystems under the ciphertext attack by performing the oracle query function only one time.

*Keyword:* public key cryptosystem, NMR ensemble search algorithm

# 1. Introduction

Recently, NMR (Nuclear Magnetic Resonance) ensemble computing [1, 2, 3, 4, 5] has attracted many researchers. An NMR ensemble computer contains many identical molecules. Each molecule, like a quantum computer, contains massive different spins which denote qubits. NMR differs from a quantum computer in that the result of a measurement is the expectation value of the observable, rather than a random eigenvalue thereof. Using this property, NMR can solve the SAT problem [2] and the search problem [1]. The search problem is to search a single object from $N=2^n$ unsorted elements. This problem can be easily solved by O(N) queries in classical computer. To reduce the number of queries, Grover [4] presented a quantum search algorithm such that the wanted object can be discovered by $O(\sqrt{N})$ queries. However, Bruschweiler's ensemble search algorithm [1] in NMR ensemble computer only requires O(logN) queries in the search of a single object in $N=2^n$ unsorted elements. In Bruschweiler's algorithm, the input state with (n+1) qubits is a mixed state whose ancillary qubit and the $i^{th}$ qubit are set to zero. Thus, the input state is mixed by N/2 pure states. The $i^{th}$ query can determine the $i^{th}$ bit value of the wanted object. After logN queries, Bruschweiler's algorithm can discover the wanted object.

In this paper, we improve Bruschweiler's algorithm such that only one query is needed for searching the wanted object. Our algorithm extends from one ancillary qubit to n ancillary qubits. Similar structure can be found in Helon and Protopopescu's algorithm [3] which uses k ancillary qubits to specify the minimal precision of summing up function values. We then construct the oracle query function g(·) satisfying g(x)=0 for all input x, except for one, say x=z, where g(z)=z. We then prepare 2n qubits, including n qubits for representing the elements and n ancillary qubits for storing the result of the oracle query function. We further obtain z by measuring these ancillary qubits.

We will use our ensemble search algorithm to attack classical public key cryptosystems. Let $E_k(m, r)$ be the encryption function, where k is a public key, m is a message, and r is a random number. We want to discover the message m from the ciphertext $C=E_k(m, r)$. Because E(·) and k are published in classical public key cryptosystems, we can generate an oracle query function h(·) satisfying

$$h(m', r') = \begin{cases} 0 & \text{if } E_k(m', r') \neq C, \\ (m', r') & \text{if } E_k(m', r') = C, \end{cases}$$

for all input (m', r'). There is only one object, say (m, r), can be discovered in decryption of C. By preparing the input with all possible states of (m', r'), we can thus use our ensemble search algorithm to find the wanted object (m, r). Obviously, we break the classical public key cryptosystems under the ciphertext attack by performing the oracle query function only one time.

The remainder of this paper is organized as follows. In Section 2, we review Bruschweiler's ensemble search algorithm. Section 3 presents a novel quantum ensemble search algorithm. Using this presented algorithm, we attack classical public key cryptosystems in Section 4. Section 5 draws the conclusions.

## 2. Review of Bruschweiler's ensemble search algorithm

Bruschweiler's ensemble search algorithm aims to search a single object in $N=2^n$ unsorted elements. Before describing Bruschweiler's algorithm, we introduce spin Liouville space which is used for NMR computation. First, the n-qubit state $|\psi\rangle = |0100...010\rangle = |\alpha\beta\alpha\alpha...\alpha\beta\alpha\rangle$, one of $2^n$ eigenstates of the Zeeman Hamiltonian, where $\alpha$ denotes spin "up" and $\beta$ denotes spin "down", can be mapped on state $s = |y\rangle\langle y| = I_1^a I_2^b I_3^a ... I_{n-1}^b I_n^a$ in spin Liouville space [5]. Here, $I_1^a I_2^b I_3^a ... I_{n-1}^b I_n^a$ is the direct product of

$$I_k^\alpha = |\alpha_k\rangle\langle\alpha_k| = \frac{1}{2}(1_k + 2I_{kz}) = \begin{pmatrix} 1 & 0 \\ 0 & 0 \end{pmatrix} \text{ and}$$

$$I_k^\beta = |\beta_k\rangle\langle\beta_k| = \frac{1}{2}(1_k - 2I_{kz}) = \begin{pmatrix} 0 & 0 \\ 0 & 1 \end{pmatrix},$$

where $k \in \{1, 2, ..., n\}$, $2I_{kz}$ is the Pauli matrix $\sigma_z$ and $1_k$ is the unity operator of the subspace of spin $I_k$. Spin $I_k$ can also be viewed as a qubit. According to $2^n$ n-qubit states $|y_0\rangle = |00...00\rangle, |y_1\rangle = |00...01\rangle, ..., |y_{2^n-1}\rangle = |11...11\rangle$, we have

$s_0 = I_1^a I_2^a ... I_{n-1}^a I_n^a, s_1 = I_1^a I_2^a ... I_{n-1}^a I_n^b, ..., s_{2^n-1} = I_1^b I_2^b ... I_{n-1}^b I_n^b$ in spin Liouville space.

Without loss of generality, we assume that the value of the element is equal to its index. For example, the third element is 3. If we want to search z, we define an oracle query function f: $\{0,1\}^n \to \{0,1\}$ that outputs 0 for all input x, f(x)=0, except for one, say x=z, where f(z)=1. In spin Liouville space, we rewrite f(·) as

$$f(s_x) = \begin{cases} 0, & \text{if } x \neq z, \\ 1, & \text{if } x = z. \end{cases}$$

We further prepare an ancillary qubit $I_0$, which is initially set to α state, to store the output of f(·). Thus, the action of f(·) can be described as a permutation with (n+1) qubits [1]. In spin Liouville space, we can construct the unitary transformation $U_f$ satisfying

$$U_f(I_0^a s) = \begin{cases} I_0^a s, & \text{if } f(s) = 0, \\ I_0^b s, & \text{if } f(s) = 1. \end{cases}$$

for the input state $I_0^a s$. Then, the evaluation of f(·) is determined by the sign of the resonance of the readout spin $I_0$. If the sign is -1 ($I_0^a$), f($s$)=0; if it is +1 ($I_0^b$), f($s$)=1. The sign can be further determined by the measurement on $I_0$. We can rewrite f($s$) as

$$f(s) = \frac{1}{2} - Tr\{U_f I_0^a s U_f^+ I_{0z}\}. \tag{1}$$

If we want to evaluate the sum of function evaluation of M states, say $\sum_{j=1}^{M} f(s_{i_j})$ for $s_{i_j}$ is a state from $\{s_0, s_1, ..., s_{2^n-1}\}$, we generally require M function evaluations in classical computer. However, using Equation (1), we have

$$\sum_{j=1}^{M} f(\boldsymbol{s}_{i_j}) = f(\sum_{j=1}^{M} \boldsymbol{s}_{i_j}) + \frac{M-1}{2},$$

where the input state $\sum_{j=1}^{M} \boldsymbol{s}_{i_j}$ is mixed by a linear combination of M states. We then require only one function evaluation to compute $\sum_{j=1}^{M} f(\boldsymbol{s}_{i_j})$. But, how to compute $\sum_{j=1}^{M} f(\boldsymbol{s}_{i_j})$ by using NMR measurement? When we measure $I_0$ of $U_f(I_0^a \sum_{j=1}^{M} \boldsymbol{s}_{i_j})$, we will obtain the sum of the sign of the resonance of the readout spin $I_0$. The sum will be either $-M$ or $-M+2$. By adding $(M-1)$ to the sum, we get the result -1 or +1. If the result is -1, $\sum_{j=1}^{M} f(\boldsymbol{s}_{i_j})=0$; if it is +1, $\sum_{j=1}^{M} f(\boldsymbol{s}_{i_j})=1$.

For simplicity, we describe Bruschweiler's algorithm by the following example. Assume that we want to find one element $z=(001)_2$ from 8 unsorted elements. We need the oracle query function $f(\cdot)$ that outputs 0 for all input x, $f(x)=0$, except for one, say x=z, where $f(z)=1$. We then construct an NMR system with four qubits, say $I_0$, $I_1$, $I_2$, and $I_3$, where $I_0$ is an ancillary qubit and the other qubits represent 8 elements. In spin Liouville space, we set the target state $I_1^a I_2^a I_3^b$ because the wanted object is z = $(001)_2$. According to the target state, we construct the unitary transformation $U_f$ satisfying $U_f(I_0^a \boldsymbol{s}_j) = I_0^a \boldsymbol{s}_j$ for all input state $I_0^a \boldsymbol{s}_j$, except for the target state $I_0^a \boldsymbol{s}_j = I_0^a I_1^a I_2^a I_3^b$, where $U_f(I_0^a I_1^a I_2^a I_3^b) = I_0^b I_1^a I_2^a I_3^b$. Let $I_0^a I_k^a$ be $I_0^a I_1^{a/b} I_2^{a/b} ... I_k^{a/b} ... I_n^{a/b}$, which is mixed by $2^{n-1}$ states. To discover the first bit value of z, we prepare the input state

$$I_0{}^a I_1{}^a = I_0{}^a I_1{}^a I_2{}^{a/b} I_3{}^{a/b} = I_0{}^a I_1{}^a I_2{}^a I_3{}^a + I_0{}^a I_1{}^a I_2{}^a I_3{}^b + I_0{}^a I_1{}^a I_2{}^b I_3{}^a + I_0{}^a I_1{}^a I_2{}^b I_3{}^b.$$

We then perform the unitary transformation $U_f$ on $I_0{}^a I_1{}^a$ and obtain

$$U_f(I_0{}^a I_1{}^a) = I_0{}^a I_1{}^a I_2{}^a I_3{}^a + I_0{}^b I_1{}^a I_2{}^a I_3{}^b + I_0{}^a I_1{}^a I_2{}^b I_3{}^a + I_0{}^a I_1{}^a I_2{}^b I_3{}^b.$$

The sign of the resonance of the readout spin $I_0$ determines the result of function f evaluation. If f(·)=0, the sign is −1 ($I_0{}^a$); otherwise, the sign is +1 ($I_0{}^b$). Thus, the second term yields the sign +1 as output on spin $I_0$ and the other terms yield -1. We can measure $I_0$ of $U_f(I_0{}^a I_1{}^a)$ and obtain the total intensity (-1)+1+(-1)+(-1) = −2. We further evaluate f($I_0{}^a I_1{}^a$)=1 by adding M-1=3, where M=$2^{3-1}$. Because f($I_0{}^a I_1{}^a$)=1, the first bit of z will be 0. Similarly, we can discover the other two bits of z by evaluating f($I_0{}^a I_2{}^a$) and f($I_0{}^a I_3{}^a$). Because $U_f(I_0{}^a I_2{}^a)$ = $I_0{}^a I_1{}^a I_2{}^a I_3{}^a + I_0{}^b I_1{}^a I_2{}^a I_3{}^b + I_0{}^a I_1{}^b I_2{}^a I_3{}^a + I_0{}^a I_1{}^b I_2{}^a I_3{}^b$, we obtain the total intensity (-1)+1+(-1)+(-1) = −2 after measuring $I_0$. We further evaluate f($I_0{}^a I_2{}^a$)=1 by adding $2^{3-1}$-1=3. The second bit of z will be 0. Next, because $U_f(I_0{}^a I_3{}^a)$ = $I_0{}^a I_1{}^a I_2{}^a I_3{}^a + I_0{}^a I_1{}^a I_2{}^b I_3{}^a + I_0{}^a I_1{}^b I_2{}^a I_3{}^a + I_0{}^a I_1{}^b I_2{}^b I_3{}^a$, we obtain the total intensity (-1)+(-1)+(-1)+(-1) = −4 after measuring $I_0$. We further evaluate f($I_0{}^a I_3{}^a$) = -1 by adding $2^{3-1}$-1=3. The third bit of z will be 1. At last, we can discover z = $(001)_2$ after three oracle queries. In general, Bruschweiler's algorithm only requires logN oracle queries to discover the wanted object from N unsorted elements.

## 3. Novel ensemble search algorithm

Assume that we want to search the object z from $N=2^n$ unsorted elements, say $a_0$, $a_1$, ..., $a_{N-1}$. Without loss of generality, let $a_i = i$. In this section, we improve Bruschweiler's algorithm by constructing new oracle query function $g(\cdot)$ satisfying $g(x)=0$ for all input x, except for one, say x=z, where $g(z)=z$. In NMR computer, we prepare n qubits $I_1$, $I_2$, ..., $I_n$, to represent the elements and n ancillary qubits $I_{01}$, $I_{02}$, ..., $I_{0n}$, to store the result of the oracle query function $g(\cdot)$. If we rewrite $z = z_1 z_2 ... z_n$, where $z_i \in \{a, b\}$, we have

$$g(s_j) = \begin{cases} 0, & \text{if } j \neq z, \\ z, & \text{if } j = z. \end{cases}$$

We can construct the unitary transformation $U_g$ satisfying

$$U_g(I_{01}{}^a I_{02}{}^a ... I_{0n}{}^a s_j) = \begin{cases} I_{01}{}^a I_{02}{}^a ... I_{0n}{}^a s_j, & \text{if } g(s_j) = 0, \\ I_{01}{}^{z_1} I_{02}{}^{z_2} ... I_{0n}{}^{z_n} s_j, & \text{if } g(s_j) = z. \end{cases}$$

Having $U_g$, we can perform the following ensemble search algorithm to discover the wanted object.

Step 1: First, we prepare 2n qubits $I_{01}$, $I_{02}$, ..., $I_{0n}$, $I_1$, $I_2$, ..., $I_n$, which are initially set to $\alpha$ states. We apply the Walsh-Hadamard Transformation H to the last n qubits and obtain the input state of $U_g$:

$$U_H(I_{01}{}^a I_{02}{}^a ... I_{0n}{}^a I_1{}^a I_2{}^a ... I_n) = I_{01}{}^a I_{02}{}^a ... I_{0n}{}^a I_1{}^{a/b} I_2{}^{a/b} ... I_n{}^{a/b} = I_{01}{}^a I_{02}{}^a ... I_{0n}{}^a.$$

Step 2: We perform the operation $U_g$ which applies $g(\cdot)$ to the last n qubits and then stores the result in the ancillary qubits. Thus, we have

$$U_g(I_{01}{}^a I_{02}{}^a \ldots I_{0n}{}^a) = I_{01}{}^{z_1} I_{02}{}^{z_2} \ldots I_{0n}{}^{z_n} s_z + \sum_{j=0, j \neq z}^{2^n-1} I_{01}{}^a I_{02}{}^a \ldots I_{0n}{}^a s_j.$$

Step 3: We perform NMR measurement on the ancillary qubits. We obtain the total intensity of $I_{01}, I_{02}, \ldots, I_{0n}$. By adding $M-1=2^n-1$, we compute the signs for $I_{01}, I_{02}, \ldots, I_{0n}$. If the sign is -1 for $I_{0i}$, we have the $i^{th}$ bit of z is 0; otherwise, if the sign is +1, the $i^{th}$ bit of z is 1. At last, we output the wanted object z according to these signs.

*Example 1:*

Assume that we want to find one element $z=(001)_2$ from 8 unsorted elements. We construct the oracle query function $g(\cdot)$ that outputs 0 for all input x, $g(x)=1$, except for one, say $x=z$, where $g(z)=z$. We then construct an NMR system with six qubits, say $I_{01}, I_{02}, I_{03}, I_1, I_2,$ and $I_3$, where $I_{01}, I_{02}, I_{03}$ are ancillary qubits and the other qubits represent 8 elements. In spin Liouville space, we set the target state $I_1{}^a I_2{}^a I_3{}^b$ because the wanted object is $z = (001)_2$. According to the target state, we construct the unitary transformation $U_g$ satisfying $U_g(I_{01}{}^a I_{02}{}^a I_{03}{}^a s_j) = I_{01}{}^a I_{02}{}^a I_{03}{}^a s_j$ for all input state $I_{01}{}^a I_{02}{}^a I_{03}{}^a s_j$, except for the target state $I_{01}{}^a I_{02}{}^a I_{03}{}^a s_j = I_{01}{}^a I_{02}{}^a I_{03}{}^a I_1{}^a I_2{}^a I_3{}^b$, where $U_g(I_{01}{}^a I_{02}{}^a I_{03}{}^a I_1{}^a I_2{}^a I_3{}^b) = I_{01}{}^a I_{02}{}^a I_{03}{}^b I_1{}^a I_2{}^a I_3{}^b$. Let the input state $I_{01}{}^a I_{02}{}^a I_{03}{}^a$ be $I_{01}{}^a I_{02}{}^a I_{03}{}^a I_1{}^{a/b} I_2{}^{a/b} I_3{}^{a/b}$, which is mixed by $2^3$ states. We follow our algorithm to discover the wanted object $z = (001)_2$.

Step 1: First, we prepare six qubits, say $I_{01}$, $I_{02}$, $I_{03}$, $I_1$, $I_2$, and $I_3$, which are set to $\alpha$ state. We then apply the Walsh-Hadamard Transformation H to $I_1$, $I_2$, $I_3$ and obtain the input state of $U_g$:

$$U_H(I_{01}{}^a I_{02}{}^a I_{03}{}^a I_1{}^a I_2{}^a I_3{}^a) = I_{01}{}^a I_{02}{}^a I_{03}{}^a.$$

Step 2: We perform the operation $U_g$ which applies $g(\cdot)$ to $I_1$, $I_2$, $I_3$ and then stores the result in the ancillary qubits. Thus, we have

$$U_g(I_{01}{}^a I_{02}{}^a I_{03}{}^a) = I_{01}{}^a I_{02}{}^a I_{03}{}^a S_0 + I_{01}{}^a I_{02}{}^a I_{03}{}^b S_1 + I_{01}{}^a I_{02}{}^a I_{03}{}^a S_2 +$$

$$I_{01}{}^a I_{02}{}^a I_{03}{}^a S_3 + I_{01}{}^a I_{02}{}^a I_{03}{}^a S_4 + I_{01}{}^a I_{02}{}^a I_{03}{}^a S_5 + I_{01}{}^a I_{02}{}^a I_{03}{}^a S_6 + I_{01}{}^a I_{02}{}^a I_{03}{}^a S_7.$$

Step 3: We perform NMR measurement on the ancillary qubits. We obtain the total intensity $-8$, $-8$ and $-6$ after measuring $I_{01}$, $I_{02}$ and $I_{03}$, respectively. By adding $2^3-1=7$, we have the sign $-1$, $-1$ and $+1$ for $I_{01}$, $I_{02}$ and $I_{03}$. At last, we output the wanted object $z = (001)_2$.

*Discussion:*

In the following, we discuss our ensemble search algorithm from three cases. First, we give the reason why oracle function $g(\cdot)$ can be implemented by a unitary operation. Second, we must modify the input size when N is not the power of 2. Third, we must face the accurate problem when the measurement error occurs.

*Case 1:*

Brüschweiler [1] showed that the oracle function $f(\cdot)$ can be implemented by the unitary operation if $f(\cdot)$ is described as a permutation function. The used oracle function $g(\cdot)$ in our algorithm is also described as a permutation function with (2n) qubits: $I_{01}, I_{02}, \ldots, I_{0n}, I_1, I_2, \ldots, I_n$. In the following, we give an example to understand that the oracle function $g(\cdot)$ can be described as a permutation function. Assume that we want to search the object $z = (10)_2$ from 4 unsorted elements. We then need two qubits, say $I_1$ and $I_2$, and two ancillary qubits, say $I_{01}$ and $I_{02}$. We construct an oracle function $g(\cdot)$ satisfying $g(0)=0$, $g(1)=0$, $g(2)=2$, and $g(3)=0$. Thus, Fig. 1 shows that $g(\cdot)$ operates on $I_1$ and $I_2$ with the output stored on $I_{01}$ and $I_{02}$. Obviously, the action of $g(\cdot)$ can be viewed as a permutation of all states spanned by $I_{01}, I_{02}, I_1$ and $I_2$. Therefore, we can implement the unitary operation $U_g$ corresponding to the oracle function $g(\cdot)$.

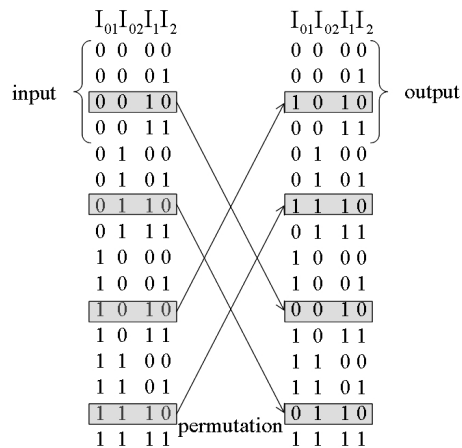

Fig. 1: Graphical representation of an oracle query function $g(\cdot)$ operating on qubits $I_1$ and $I_2$ as a permutation using two ancillary qubits $I_{01}$, $I_{02}$ with the output stored on $I_{01}$ and $I_{02}$.

*Case 2:*

If N is not the power of 2, say $2^{n-1} < N < 2^n$, we add extra $2^n - N$ elements which are different from z. We then prepare n qubits to represent these elements. Thus, our algorithm can discover the wanted object from $2^n$ elements.

*Case 3:*

The measurement sensitivity for the ancillary qubits is a limitation on our algorithm. For ith ancillary qubit, only one out of N inputs yields $I_{0i}^{b}$. These n ancillary qubits have the same measurement sensitivity because the measurement on ancillary qubits cannot influence the input states. Thus, our algorithm and Bruschweiler's [3] have the same measurement sensitivity. Similar analysis can be found in Helon and Protopopescu's algorithm [3]. Let $\delta$ be the measurement error for a single experimental trial. Assume that $\delta = \frac{1}{2^k}$. If k≥ n, the adequate measurement sensitivities give us a correct measured value. Otherwise, we may obtain a wrong measured value. However, the measurement error can be reduced by repeating our algorithm a number of times. The measurement error scales inversely with the square-root of the number of experimental trials [1, 7]. Assume that $N_\delta$ is the number of experimental trials. Then, if we want to obtain the correct measured value, $N_\delta$ must satisfy $\frac{1}{2^k} \times \frac{1}{\sqrt{N_\delta}} < \frac{1}{2^n}$. It implies that $N_\delta > 2^{2(n-k)}$. We thus require $O(2^{2(n-k)})$ oracle queries to find the wanted element. When $k \approx n$, our algorithm only requires $O(1)$ oracle query. But, when $k \ll n$, our algorithm requires $O(2^{2n})$ oracle queries.

Bruschweiler [1] also showed that his algorithm is more efficient than Grover's algorithm [4] for databases of size N satisfying

$$N\sqrt{N}\log N < \boldsymbol{d}^{-2}.$$

Similarly, our algorithm is more efficient than Grover's algorithm for databases of size N satisfying

$$N\sqrt{N} < \boldsymbol{d}^{-2}.$$

With current NMR spectrometer technology, the measurement error is at least $10^{-7}$ [7]. Thus, our algorithm efficiently searches a single object from $N \leq 2^{30}$ unsorted elements.

## 4. Attacks of classical public key cryptosystems

In this section, we will use our presented ensemble search algorithm to attack classical public key cryptosystems. Let $E_k(m, r)$ be the encryption function from $\{0, 1\}^{n_1+n_2}$ to $\{0, 1\}^n$, where k is a public key, m is a message with $n_1$ bits, and r is a random number with $n_2$ bits. Given the ciphertext $C=E_k(m, r)$ with n bits, we want to discover the message m by using our presented ensemble search algorithm. Because $E_k(\cdot)$ and k are published in classical public key cryptosystems, we can generate an oracle query function $h(\cdot)$ satisfying

$$h(m', r') = \begin{cases} 0 & \text{if } E_k(m', r') \neq C, \\ (m', r') & \text{if } E_k(m', r') = C, \end{cases}$$

for all input (m', r'). There is only one object, say (m, r), can be discovered in decryption of C. In NMR computer, we prepare $n_1$ qubits $I_{11}, I_{12}, \ldots, I_{1n_1}$, to represent the messages, $n_2$ qubits $I_{21}, I_{22}, \ldots, I_{2n_2}$, to represent random numbers and $(n_1 + n_2)$ ancillary qubits $I_{01}, I_{02}, \ldots, I_{0(n_1+n_2)}$, to store the result of the oracle query function $h(\cdot)$.

Let

$$s_{10} = I_{11}{}^a I_{12}{}^a \ldots I_{1(n_1-1)}{}^a I_{1n_1}{}^a, s_{11} = I_{11}{}^a I_{12}{}^a \ldots I_{1(n_1-1)}{}^a I_{1n_1}{}^b, \ldots, s_{1(2^{n_1}-1)} = I_{11}{}^b I_{12}{}^b \ldots I_{1(n_1-1)}{}^b I_{1n_1}{}^b,$$

$$s_{20} = I_{21}{}^a I_{22}{}^a \ldots I_{2(n_2-1)}{}^a I_{2n_2}{}^a, s_{21} = I_{21}{}^a I_{22}{}^a \ldots I_{2(n_2-1)}{}^a I_{2n_2}{}^b, \ldots, s_{2(2^{n_2}-1)} = I_{21}{}^b I_{22}{}^b \ldots I_{2(n_2-1)}{}^b I_{2n_2}{}^b.$$

If we rewrite $m = m_1 m_2 \ldots m_{n_1}$ and $r = r_1 r_2 \ldots r_{n_2}$, where $m_i, r_i \in \{a, b\}$, we have

$$h(s_{1j_1}, s_{2j_2}) = \begin{cases} 0, & \text{if } E_k(j_1, j_2) \neq C, \\ (j_1, j_2), & \text{if } E_k(j_1, j_2) = C. \end{cases}$$

We can construct the unitary transformation $U_h$ which applies $h(\cdot)$ to $I_{11}, I_{12}, \ldots, I_{1n_1}$, $I_{21}, I_{22}, \ldots, I_{2n_2}$, and stores the result in $I_{01}, I_{02}, \ldots, I_{0(n_1+n_2)}$. Thus, we have

$$U_h(I_{01}{}^a I_{02}{}^a \ldots I_{0(n_1+n_2)}{}^a s_{1j_1} s_{2j_2}) = \begin{cases} I_{01}{}^a I_{02}{}^a \ldots I_{0(n_1+n_2)}{}^a s_{1j_1} s_{2j_2}, & \text{if } E_k(j_1, j_2) \neq C, \\ I_{01}{}^{m_1} I_{02}{}^{m_2} \ldots I_{0n_1}{}^{m_{n_1}} I_{0(n_1+1)}{}^{r_1} \ldots I_{0(n_1+n_2)}{}^{r_{n_2}} s_{1j_1} s_{2j_2}, & \text{if } E_k(j_1, j_2) = C. \end{cases}$$

Having $U_h$, we can perform the following ensemble search algorithm to discover the message m.

Step 1: First, we prepare $2(n_1 + n_2)$ qubits $I_{01}, I_{02}, \ldots, I_{0(n_1+n_2)}, I_{11}, I_{12}, \ldots, I_{1n_1}, I_{21}, I_{22}, \ldots, I_{2n_2}$, which are initially set to $\alpha$ states. We apply the Walsh-Hadamard Transformation H to the last $(n_1 + n_2)$ qubits $I_{11}, I_{12}, \ldots, I_{1n_1}, I_{21}, I_{22}, \ldots, I_{2n_2}$, and obtain the input state of $U_h$:

$$U_H(I_{01}{}^a I_{02}{}^a ... I_{0(n_1+n_2)}{}^a I_{11}{}^a I_{12}{}^a ... I_{1n_1}{}^a I_{21}{}^a I_{22}{}^a ... I_{2n_2}{}^a) = I_{01}{}^a I_{02}{}^a ... I_{0(n_1+n_2)}{}^a.$$

Step 2: We perform the operation $U_h$ which applies $h(\cdot)$ to the last $(n_1+n_2)$ qubits and stores the result in the ancillary qubits. Thus, we have

$$U_h(I_{01}{}^a I_{02}{}^a ... I_{0(n_1+n_2)}{}^a) =$$
$$I_{01}{}^{m_1} I_{02}{}^{m_2} ... I_{0n_1}{}^{m_{n_1}} I_{0(n_1+1)}{}^{r_1} ... I_{0(n_1+n_2)}{}^{r_{n_2}} S_{1m} S_{2r} + \sum_{\substack{j_1=0 \\ j_1 \neq m}}^{2^{n_1}-1} \sum_{\substack{j_2=0 \\ j_2 \neq r}}^{2^{n_2}-1} I_{01}{}^a I_{02}{}^a ... I_{0(n_1+n_2)}{}^a S_{1j_1} S_{2j_2}$$

Step 3: We perform NMR measurement on the ancillary qubits. We obtain the total intensity of $I_{01}, I_{02}, ..., I_{0(n_1+n_2)}$. By adding $M-1=2^{(n_1+n_2)}-1$, we compute the signs for $I_{01}, I_{02}, ..., I_{0n_1}$. If the sign is $-1$ for $I_{0i}$, the $i^{th}$ bit of m is 0; otherwise, if the sign is $+1$, the $i^{th}$ bit of m is 1. At last, we output the message m according to these signs.

In the following, we use our algorithm to attack two cryptosystems: RSA [8] and McEliece's [6] cryptosystems.

*Example 2:* **Attack on RSA [8]**

In this example, we use our algorithm to attack RSA cryptosystem. Let (e, N) be the RSA public key. The RSA encryption function is $E_{(e, N)}(m, r) = m^e \mod N$. Obviously, the RSA encryption function doesn't require random numbers. For simplicity, we choose $(e, N) = (7, 15)$ to explain our attack. Given the ciphertext $C = m^e \mod N = 2$, which is generated by the message $m=8$, we construct the oracle query

function h(·) satisfying

$$h(s_{1j_1}) = \begin{cases} 0, & \text{if } E_{(7,15)}(j_1) \neq 2, \\ j_1, & \text{if } E_{(7,15)}(j_1) = 2. \end{cases}$$

Because N is of 4 bits and $m = m_1 m_2 m_3 m_4 = \alpha\beta\alpha\alpha$, we use 8 qubits $I_{01}$, $I_{02}$, $I_{03}$, $I_{04}$, $I_{11}$, $I_{12}$, $I_{13}$, $I_{14}$, to construct the unitary transformation $U_h$ satisfying

$$U_h(I_{01}{}^a I_{02}{}^a I_{03}{}^a I_{04}{}^a s_{1j_1}) = \begin{cases} I_{01}{}^a I_{02}{}^a I_{03}{}^a I_{04}{}^a s_{1j_1}, & \text{if } E_{(7,15)}(j_1) \neq 2, \\ I_{01}{}^{m_1} I_{02}{}^{m_2} I_{03}{}^{m_3} I_{04}{}^{m_4} s_{1j_1}, & \text{if } E_{(7,15)}(j_1) = 2. \end{cases}$$

Having $U_h$, we perform the following steps to discover message m.

Step 1: We prepare 8 qubits $I_{01}$, $I_{02}$, $I_{03}$, $I_{04}$, $I_{11}$, $I_{12}$, $I_{13}$, $I_{14}$, which are initially set to $\alpha$ states. We apply the Walsh-Hadamard Transformation H to the last 4 qubits $I_{11}$, $I_{12}$, $I_{13}$, $I_{14}$, and obtain the input state of $U_h$:

$$U_H(I_{01}{}^a I_{02}{}^a I_{03}{}^a I_{04}{}^a I_{11}{}^a I_{12}{}^a I_{13}{}^a I_{14}{}^a) = I_{01}{}^a I_{02}{}^a I_{03}{}^a I_{04}{}^a.$$

Step 2: We perform the operation $U_h$ which applies h(·) to the last 4 qubits and stores the result in the ancillary qubits. Thus, we have

$$U_h(I_{01}{}^a I_{02}{}^a I_{03}{}^a I_{04}{}^a) = I_{01}{}^b I_{02}{}^a I_{03}{}^a I_{04}{}^a s_{18} + \sum_{\substack{j_1=0 \\ j_1 \neq 8}}^{15} I_{01}{}^a I_{02}{}^a I_{03}{}^a I_{04}{}^a s_{1j}$$

Step 3: We perform NMR measurement on the ancillary qubits. We obtain the total intensity $-14$, $-16$, $-16$ and $-16$ after measuring $I_{01}$, $I_{02}$, $I_{03}$ and $I_{04}$, respectively. By adding $2^4 - 1 = 15$, we have the sign $+1$, $-1$, $-1$ and $-1$ for $I_{01}$, $I_{02}$, $I_{03}$ and $I_{04}$. At last, we output the wanted object $m = (1000)_2 = 8$.

*Example 3:* **Attack on McEliece's public key cryptosystem [6]**

In this example, we use our algorithm to attack McEliece's public key cryptosystem. Let G' be the k×n generator matrix of a Goppa codes that corrects up to t errors. Let the public key be G=SG'P, where S is an inversible k×k matrix and P is a random n×n permutation matrix. The encryption function is $E_G(m, r) = mG + r$, where m is a message matrix with dimension 1×k and r is a random error vector of size n that has at most t' entries for t' ≤ t. Assume that t' is known. We can obtain the domain of r, say $r_0^{'}$, $r_1^{'}$, …, $r_d^{'}$. For simplicity, we choose k = 2, n=7, t=2 and G=$\begin{bmatrix} 1101 0101 \\ 1000 1111 \end{bmatrix}$. We can deduce that d ≈ 28 from t=2. We can rearrange these random numbers with 5 bits by $s_{20} = r_0^{'}, s_{21} = r_1^{'},...,s_{2d} = r_d^{'}$. Assume that the ciphertext C = mG + r =$\begin{bmatrix} 0101 0011 \end{bmatrix}$, which is generated by the message m=[1 1] and r = $\begin{bmatrix} 0000 1001 \end{bmatrix}$. Assume that r = $r_{10}^{'}$. In this example, we must transfer the matrix into the binary number. Thus we have G=$(1101 0101 1000 1111)_2$, C = $(0101 0011)_2$. According to C = $(0101 0011)_2$, we construct the oracle query function h(·) satisfying

$$h(s_{1j_1}, s_{2j_2}) = \begin{cases} 0, & \text{if } E_G(j_1, j_2) \neq (0101 0011)_2, \\ (j_1, j_2), & \text{if } E_G(j_1, j_2) = (0101 0011)_2. \end{cases}$$

Because m and r are respectively represented by 2 bits and 5 bits, we can construct the unitary transformation $U_h$, which applies h(·) to $I_{11}$, $I_{12}$, $I_{21}$, $I_{22}$, …, $I_{25}$, and stores the result in $I_{01}$, $I_{02}$, …, $I_{07}$. Thus, we have

$$U_h(I_{01}{}^a I_{02}{}^a ... I_{07}{}^a \mathcal{S}_{1j_1}\mathcal{S}_{2j_2}) = \begin{cases} I_{01}{}^a I_{02}{}^a ... I_{07}{}^a \mathcal{S}_{1j_1}\mathcal{S}_{2j_2}, & \text{if } E_G(j_1, j_2) \neq (01010011)_2, \\ I_{01}{}^{m_1} I_{02}{}^{m_2} I_{03}{}^{r_1} ... I_{07}{}^{r_7} \mathcal{S}_{1j_1}\mathcal{S}_{2j_2}, & \text{if } E_G(j_1, j_2) = (01010011)_2. \end{cases}$$

Having $U_h$, we perform the following steps to discover message m.

Step 1: We prepare 14 qubits $I_{01}, I_{02}, ..., I_{07}, I_{11}, I_{12}, I_{21}, I_{22}, ..., I_{25}$ which are initially set to $\alpha$ states. We apply the Walsh-Hadamard Transformation H to the last 7 qubits $I_{11}, I_{12}, I_{21}, I_{22}, ..., I_{25}$, and obtain the input state of $U_h$:

$$U_H(I_{01}{}^a I_{02}{}^a ... I_{07}{}^a I_{11}{}^a I_{12}{}^a I_{21}{}^a I_{22}{}^a ... I_{25}{}^a) = I_{01}{}^a I_{02}{}^a ... I_{07}{}^a.$$

Step 2: We perform the operation $U_h$ which applies $h(\cdot)$ to the last 7 qubits and then store the result in the ancillary qubits. Thus, we have

$$U_h(I_{01}{}^a I_{02}{}^a ... I_{07}{}^a) = I_{01}{}^b I_{02}{}^b I_{03}{}^a I_{04}{}^b I_{05}{}^a I_{06}{}^b I_{07}{}^a \mathcal{S}_{13}\mathcal{S}_{2(10)} + \sum_{\substack{j_1=0 \\ (j_1 \neq 3 \text{ and } }}^{3} \sum_{\substack{j_2=0 \\ j_2 \neq 10)}}^{31} I_{01}{}^a I_{02}{}^a ... I_{07}{}^a \mathcal{S}_{1j_1}\mathcal{S}_{2j_2}.$$

Step 3: We perform NMR measurement on the ancillary qubits. We obtain the total intensity –126, -126 after measuring $I_{01}, I_{02}$, respectively. By adding $2^7-1=127$, we have the sign +1, +1 for $I_{01}, I_{02}$. At last, we output the wanted object m= $(11)_2$.

*Discussion:*

Section 3 gives us that our algorithm can efficiently solve the searching problem for the database of the size $N \leq 2^{30}$. Therefore, we only attack the RSA cryptosystem with modulus N of less than 30 bits. Similarly, we efficiently break McEliece's public key cryptosystem when k+d ≤ 30.

## 5. Conclusion

This paper has presented the new oracle query function g(·) to improve Bruschweiler's algorithm such that only one query is needed for searching the single object z from $N=2^n$ unsorted elements. The new oracle query function g(·) satisfies g(x)=0 for all input x, except for one, say x=z, where g(z)=z. To store z, our algorithm extends from one ancillary qubit to n ancillary qubits. We then measure these ancillary qubits to discover z. According to our analysis, our algorithm can efficiently discover the wanted object if $N \leq 2^{30}$. We further use our ensemble search algorithm to attack classical public key cryptosystems. Given the ciphertext C=$E_k$(m, r), we can construct an oracle query function h(·) satisfying h(m', r')=0 if $E_k$(m', r')≠C and h(m', r')= (m', r') if $E_k$(m', r')=C. There is only one object, say (m, r), can be discovered in decryption of C. By preparing the input with all possible states of (m', r'), we can thus use our ensemble search algorithm to find the wanted object (m, r). With current NMR spectrometer technology, we break the RSA cryptosystem with up to 30-bit modulus.